\documentstyle[prl,aps]{revtex}
\tighten
\begin{document}
\draft

\title{
Coulomb Blockade without Tunnel Junctions
       }

\author{ Yuli V. Nazarov}
\address{
Faculty of Applied Physics and DIMES, Delft University of Technology,
Lorentzweg 1, 2628 CJ Delft,\\
The Netherlands
	}
\maketitle
\begin{abstract}
We find that tunnel junctions are not needed to provide single 
electron effects in a metallic island.
Eventually the tunnel junction may be replaced by an arbitrary scatterer.    
It is important that even a {\it diffusive} scatterer provides
a sufficient isolation for single electron  effects to persist.
To formulate this in exact terms, we derive and analyze 
the effective action that describes an arbitrary scatterer.
We also consider the fluctuations of the effective charging energy.
\end{abstract} 


It is well-known that electric charge of an isolated piece of conducting
material can only take discrete values corresponding to the integer number of
electrons in there. This property persists if this isolated piece, the island,
is connected to electron reservoir by means of a resistive tunnel junction
It is the recognition of this mere fact that lead to outburst
of the entire field of single electron phenomena.\cite{SingleElectronTunneling}   

The single electron effects are  best visible provided
the conductivity of the tunnel junction is much smaller than
the conductance quantum $G_Q \equiv e^2/2 \pi \hbar$. The ground
state energy as a function of induced charge $q$ is given by
minimization of Coulomb energy, $E_C(n+q)^2$, with respect to
discrete charge $n$. The result is periodic in $q$ with a period
$e$.\cite{SingleElectronTunneling}
The analysis of the reverse case, $G \gg G_Q$, 
requires advanced  theoretical methods. 
\cite{PanZa,Grabert,Willy} 
Despite the partial controversy in results, all authors agree that 
in this case the ground state
energy retain the periodic $q$ dependence, that manifests the Coulomb blockade.
The effective charging energy, $\tilde E_C$, that is, the $q$-dependent part of
the ground state energy, is suppressed by a factor of $\exp(-G/2 G_Q)$ in comparison
with $E_C$.

Still the analysis has been restricted to tunnel junctions.
Next step has been made in \cite{Matveev,Flensberg} where  Coulomb
blockade has been studied in the situation where the isolation is provided
by a quantum point contact with almost perfect transparency.
It has been shown that the charge quantization survives. Albeit the charging
energy is strongly suppressed vanishing to zero at perfect transmission.

In this paper, we construct a general theory of Coulomb blockade that
can embrace tunnel junctions, quantum point contacts, diffusive conductors
and eventually any type of scattering. 

The results are as follows. Charging energy vanishes only for perfect point
contacts. For very wide class of conductors that have conductivity $G \gg G_Q$, 
the charging energy is exponentially suppressed, 
$\ln (\tilde E_C/E_C) \propto -\alpha G/G_Q$, $\alpha$ being
dimensionless coefficient depending on the type of the conductor. 
For disordered conductors, for instance, diffusive ones, the charging energy 
strongly fluctuates. 
This happens even if the fluctuations of the conductance are small.

The most equivalent mathematical framework to describe the charging effects
in full has been proposed  by Sch\"{o}n and Zaikin.\cite{SZReview}  
 They present the partition
function of the system in the form of the path integral over the field
$\varphi(\tau)$($\beta=\hbar/T$),
\begin{equation}
Z=\int \prod_{\tau} d \varphi(\tau) \exp(-{\cal L}_{sc}[\varphi(\tau)] +
\int_0^\beta d\tau (- \frac{{\dot{\varphi}(\tau)}^2}{2 E_C}
 -i \frac{q \dot{\varphi}(\tau)}{e}) 
\end{equation}
The form of the last term presumes that the partition function
can be presented as a sum over topological sectors that are labelled by
an integer $W = \int_0^\beta d\tau \dot{\varphi}(\tau)$
winding number of $\varphi(\tau)$,
\begin{equation}
Z=\sum_W Z_W \exp( -i 2 \pi W q/e )
\end{equation}
Here $Z_W$ does not depend on $q$. 

The action ${\cal L}_{sc}$ describes tunnel junction. It can be evaluated 
with using Tunneling Hamiltonian method that gives 
\begin{equation}
-{\cal L}_{sc} = \frac{G_T}{2 \beta^2 G_Q}
\int_0^\beta d\tau \int_0^\beta d\tau' \frac{\sin^2((\varphi(\tau)-\varphi(\tau'))/2)}
{ \sin^2(\pi (\tau-\tau')/\beta)}.
\label{tunneljunction} 
\end{equation}
This form shows that the tunnel junction is quite different form a
linear resistor which is described by a form
\begin{equation}
-{\cal L}_{sc} = \frac{G}{ 8 \beta^2 G_Q}
\int_0^\beta d\tau \int_0^\beta d\tau' \frac{(\varphi(\tau)-\varphi(\tau'))^2}
{ \sin^2(\pi (\tau-\tau')/\beta)}
\label{linear}
\end{equation}
which is bilinear in $\varphi$. It has been frequently assumed that a coherent
diffusive conductor can be described by (\ref{linear}) and consequently exhibits
no charging effects. We show below that it is not so. However, the
relation (\ref{linear}) holds for an arbitrary conductor in the limit of
small $\varphi$.

We sketch the derivation of ${\cal L}_{sc}$ for an arbitrary conductor.
Our basic assumptions are: (i) no inelastic scattering occurs in the conductor,
(ii) the conductor is sufficiently short, $E_C\tau_{trav}/\hbar \ll 1$,
$\tau_{trav}$ being typical traversal time throught the conductor,
(iii) the island is sufficiently large, so that $E_C$ greatly exceeds
the average level spacing in the island.
Two first assumptions allow us to characterize the conductor by an elastic
scattering matrix disregarding retardation effects. The third assumption
allows us to disregard coherence between the electrons transmitted to and coming from
the island, so that it can be regarded as an electron reservoir.

These assumptions correspond to a fermionic action of the following form:
\begin{eqnarray}
-{\cal L}= \int_0^\beta d\tau ( 
\sum_{n=1}^N \int_{-\infty}^0 d x
\chi_n^{\dagger}(x,\tau) \left[\partial_{\tau} +i v_n \partial_{x} \right]\chi_n(x,\tau)
+\psi_n^{\dagger}(x,\tau) \left[\partial_{\tau} -i v_n \partial_{x} \right]\psi_n(x,\tau) \\ \nonumber
+\sum_{m=1}^M \int_{0}^{\infty} d x
\chi_n^{\dagger}(x,\tau) \left[\partial_{\tau} -i v_n \partial_{x} \right]\chi_n(x,\tau)
+\psi_n^{\dagger}(x,\tau) \left[\partial_{\tau} +i v_n \partial_{x} \right]\psi_n(x,\tau) \\ \nonumber
+ E_C (Q(\tau)-q/e)^2)
\end{eqnarray}
Here the island is on the right ($x>0$), $n$ and $m$ label transport
channels in the island and the reservoir respectively. $\chi$ stands for
fermion fields coming to the scatterer and $\psi$ stands for outgoing
modes. The scatterer is completely characterized by the scattering
matrix $\hat S_{kl}$ that set a boundary condition for $\psi$ and $\chi$
\begin{equation}
\psi_k(0) = \sum_l S_{kl} \chi_l(0).
\end{equation}
Here $k$, $l$ label modes on both sides of the scatterer. 

The charge in the island is given by
\begin{equation}
Q(\tau) = \sum_{m=1}^M \int_{0}^{\infty} dx 
\left( \chi_m^{\dagger}(x,\tau)  \chi_m(x,\tau)
+\psi_m^{\dagger}(x,\tau)  \psi_m(x,\tau) \right)
\end{equation}
To proceed, we perform a Hubbard-Stratanovitch transform on interaction term
introducing a new variable $\varphi(\tau)$,
\begin{equation}
E_C (Q(\tau)-q/e)^2) \rightarrow - \frac{\dot{\phi(\tau)}^2}{2 E_C}
+ iQ(\tau) \dot{\phi(\tau)} -i q/e \dot{\phi(\tau)}
\end{equation}  
  
The resulting action is quadratic in fermions so that they can be
integrated out and the action can be represented as a functional of
$\varphi(\tau)$. It is constructive to do this calculating 
Green functions of the fermions in the presence of field $\phi$
and scattering potential. Special attention shall be given to the fact
that in one dimension the Green function $G(x,\tau,x',\tau';[\varphi(\tau)])$ is not
continuous at coinciding arguments. So that the problem shall be regularized
by letting scattering to occur in small but finite region of space.
We present the action as a series in $\Delta G =G_0(x,\tau,x',\tau';[\varphi(\tau)=0])-G_0(x,\tau,x',\tau';[\varphi(\tau)])$
where $G_0$ refers to Green functions of disconnected island and
reservoir, since $\Delta G$ is continuous. Then the action can be reduced
to the trace of logarithm of an operator, 
\begin{equation}
- {\cal L}_{sc} = {\rm Tr}_{n,\tau} \ln (1-
(1-\hat r)\hat\theta^+ \exp(-i\varphi) \hat\theta^- \exp(i \varphi) 
-(1-\hat r^{+})\hat\theta^- \exp(-i\varphi) \hat\theta^+ \exp(i
\varphi))
\label{action}
\end{equation}

The operator is defined in a Gilbert space which is the direct product
of the Gilbert space of the transport channels in the island
and the Gilbert space of fermion functions in imaginary time.
The operator $\hat r$ is the reflection matrix in the island.
The operators $\hat \theta^{\pm}$ are projection operators on positive
(negative) Matsubara frequencies, $\hat \theta^{\pm} \equiv \theta(\pm \epsilon_n)$.
Although $\hat \theta^{\pm}$ obviously commute with $\hat r,\hat r^{+}$, they
do not commute with multiplication by $\exp(\pm i \varphi(\tau))$. This generates
a complicated operator algebra that makes a complicated problem even the
evaluation of ${\cal L}_{sc}$.

The expression (\ref{action}) can be explicitly evaluated in two limits:
$\hat r \rightarrow 1$ and $\phi \rightarrow 1$. There, we successfully reproduce
Eq. \ref{tunneljunction} for tunnel junctions and Eq. \ref{linear} for an
arbitrary scatterer in linear regime.

Fortunately, a very important part of analysis of the action (\ref{action})
can be done exactly. We are able to find the minima of (\ref{action}) in each
topological sector and thus give a quantitative estimate of effective
Coulomb energy in the limit $G \gg G_Q$. We consider the configurations of 
$\varphi(\tau)$ of the following form:  
\begin{equation}
\exp(i \varphi) = \prod_{i=1}^{N} \frac{u - z_i}{1-u z^*_i}. 
\label{soliton}
\end{equation}
Here $u \equiv \exp(i 2 \pi \tau/\beta)$, $z_i$ are complex parameters.
$z_i$ can be viewed as coordinates of $N$ (anti)solitons in the plane of
complex $u$ (see Fig.2). If $|z_i| \rightarrow 1$, these configurations correspond
to sets of Korshunov's solitons.\cite{Korshunov}
 We are interested in configurations where all solitons are of the same sign. 
In this case either $|z_i| <1$ for all $i$ or $|z_i| >1$ so that $\exp(i \varphi)$
is an analytical function of $u$ either within or beyond the unitary circle. The
winding number $W =\pm N$. Using the methods of analytical function theory
we show that these configurations indeed minimize the action in the
corresponding topological sector. The minimum does not depend on $z_i$
and equals 
\begin{equation}
-{\cal L}_W = \frac{1}{2} \ln {\rm det}(\hat r \hat r^+) |W| + \frac{1}{2} \ln {\rm det}(\hat r/\hat r^+) W. 
\label{mimimum}
\end{equation}
The second term is imaginary and can be viewed as a trivial shift of
induced charge $q$: $q \rightarrow q+ i \ln {\rm det} (\hat r/\hat r+)$.
The first term is of importance since it describes the suppression
of statistical weight of topogical sectors with $W \ne 0$ in comparison
with the trivial sector. It has been shown in \cite{PanZa,Grabert} that
the suppression of these statistical weights leads to suppression of effective
charging energy. This allows us to write down a simple formula for effective
charging energy 
\begin{equation}
\tilde E_C \propto E_C \prod_n R^{1/2}_n
\label{main}
\end{equation}
where $R_n$ are eigenvalues of the reflection matrix $\hat r \hat
r^{+}$. This formula is valid provided the suppression is big.
 Similar relation has been obtained by Flensberg \cite{Flensberg}
in a much more restrictive framework. In the limit of almost perfect
transmission, $R \rightarrow 0$, we reproduce the results of Matveev 
\cite{Matveev}. 

Recent theoretical advantages 
allow to characterize $R_n$ of a scatterer/conductor of virtually any
type (see \cite{CarloReview} for review). This makes the relation (\ref{main})
easy to  use for concrete examples. From now on, we will concentrate on diffusive
conductor in the limit $G \gg G_Q$. It is a disordered conductor, so that it is
characterized by distribution of $R_n$, or transmissions $T_n =1-R_n$. 
It has been shown in \cite{Nazarov_transmission} that the transmission
distribution of a diffusive conductor depends only on its conductance, 
$\rho(T)= G/(2 G_Q T \sqrt{1-T})$.
We average logarithm of (\ref{main}) with this distribution to obtain
\begin{equation}
\tilde E_C/E_C \propto \exp (-\frac{\pi^2 G}{8 G_Q}).
\label{supp_diff}
\end{equation} 
This is the main result of this work. The diffusive scatterer of
the same resistance as a tunnel junction suppresses Coulomb blockade
much more efficiently. To give some numbers, let us choose $1/G= {\rm 4}\  {\rm kOhm}$.
In this case, suppression factor is about 25 for a tunnel junction and
almost 3000 for a diffusive conductor.

Below we consider the fluctuations of $\tilde E_C$ and the effect of
weak localization. To make a qualitative estimation, we note that the
fluctuation of $G$ is of the order of $G_Q$. The weak localization
correction is of the same scale. Therefore, the fluctuations of
an exponential like (\ref{supp_diff}) must be of the order of its
average value. The same should hold for the weak localization effect.
It is remarkable that quantitative consideration gives even {\it bigger} 
values.

This quantitative treatment can be performed along the lines of
\cite{Nazarov_fluctuations} and \cite{Nazarov_wl}. There are 
formulas that can be directly applied to the quantity of interest
$ \ln(\tilde E_C/E_C) = \sum_n \ln{R_n}$. It appears that both
the fluctuation and the localization effect are dominated by a contribution
of the {\it universal} cooperon-diffusion mode, the one which provides
Wigner-Dyson statistics of closely spaced transmission eigenvalues.
\cite{Nazarov_fluctuations} The contribution of this mode logarithmically 
diverges at very small $R$ and shall be cut off at $R \simeq G_Q/G$, average
value of transmission spacing. For pure statistical ensembles, 
the fluctuation is given by 
\begin{equation}
<< \ln^2(\tilde E_C/E_C)>> = \frac{N_{dc}}{4} \ln (G/G_Q)
\end{equation}
where $N_{dc}$ is the number of massless cooperon and diffusion modes.
It ranges from 1 to 8. The weak localization correction is 
\begin{equation}
<\ln(\tilde E_C/E_C)>_{wl} = -\frac{N_{wl}}{4} \ln(G/G_Q)
\end{equation}
where $N_{wl}$ is 2, 0, -1 for simplectic, unitary and orthogonal  ensemble respectively.

Experimentally, the fluctuation and weak localization effect are
identified by their magnetic field dependence. Following
\cite{Nazarov_fluctuations} we introduce dimensionless parameters
$\eta_H$,$\eta_{SO}$ to characterize magnetic field and spin-orbit
interaction. We disregard influence of magnetic field on spin.
Correlator of two $\tilde E_C$ taken at different values
of magnetic field reads
\begin{equation}
<< \ln(\tilde E_C(\eta_H)/E_C)\ln(\tilde E_C(\eta'_H)/E_C)>>=
-\frac{1}{4} \ln(|\eta_H^2 -{\eta'_H}^2|^4 
((\eta_H+\eta'_H)^2+\eta_{SO}^2)^3 ((\eta_H-\eta'_H)^2+\eta_{SO}^2)^3).
\end{equation}
This equation is valid provided $G/G_Q \ll
|\eta_H-\eta'_H|,\eta_H,\eta'_H,\eta_{SO} \ll 1$. It shows that the actual value
of $\tilde E_C$ can be changed by an order of magnitude by fairly small
change of magnetic field $\delta \eta_H \simeq G_Q/G$. At a bigger $\delta
\eta_H$ effective charging energy exhibits power-law correlations.

Magnetic field dependence of the weak localization correction is given
by
\begin{equation}
<\ln(\tilde E_C/E_C)>_{wl} = \frac{1}{4} ( 3 \ln(\eta_H^2 + \eta_{SO}^2)
-\ln( \eta_H^2)).
\end{equation}

To conclude, we have shown that the isolation required for discrete charge
effects can be provided by any constriction which is not ideally
ballistic. We have discussed suppression of the effective charging
energy by a diffusive scatterer and found gigantic fluctuations of
this quantity. 

I am indebted to G. W. E. Bauer, P. W. Brouwer, S. E. Korshunov, Y. Gefen
and many others for very instructive discussions of the results.
This work is a part of the research programme of the "Stichting voor
Fundamenteel Onderzoek der Materie"~(FOM) and I acknowledge the financial
support from the "Nederlandse Organisatie voor Wetenschappelijk Onderzoek"
~(NWO).

\end{document}